# Investigation of Temperature Sensitivity of a Geometric Anti-spring based MEMS Gravimeter

Vinod Belwanshi[1,a)], Abhinav Prasad[1], Karl Toland[1], Richard Middlemiss[1], Douglas Paul[2], Giles Hammond[1]

**AFFILIATIONS**

[1] Institute for Gravitational Research, School of Physics and Astronomy, University of Glasgow, Glasgow, UK G12 8QQ
[2] James Watt School of Engineering, School of Engineering, University of Glasgow, Glasgow, UK G12 8LT

[a)]**Author to whom correspondence should be addressed:** vinod.belwanshi@glasgow.ac.uk

**ABSTRACT**

This paper describes a temperature sensitivity or thermal sag measurement of a geometric anti-spring based micro-electromechanical system (MEMS) gravimeter (Wee-g). The Wee-g MEMS gravimeter is currently fabricated on a (100) silicon wafer using standard micro-nano fabrication techniques. The thermal behavior of silicon indicates that the Young's modulus of silicon decreases with an increase in temperature (~63 ppm/K). This leads to a softening of the silicon material resulting in the proof mass (PM) displacing (or sagging) under the influence of an increasing temperature. It results in the change on the measured gravity and it is expressed as a temperature sensitivity in terms of change in gravity per degree temperature. The temperature sensitivity for the silicon based MEMS gravimeter is found to be 60.14 - 64.87 µGal/mK, 61.76 µGal/mK and 62.76 µGal/mK for experimental, finite element analysis (FEA) simulation and analytical calculations respectively. It suggests that temperature sensitivity is depended on material properties used to fabricate the MEMS devices. In this paper the experimental measurements of thermal sag are presented, along with analytical calculations and simulations of the effect using FEA (finite element analysis). The bespoke optical measurement system to quantify the thermal sag is also described. The results presented are an essential step towards the development of temperature insensitive MEMS gravimeters.

## I. INTRODUCTION

Microelectromechanical system (MEMS) based silicon sensors have gained popularity in a variety of application areas and gravimetry is amongst them. The MEMS gravimeter, developed at Glasgow, utilizes geometrical anti-spring (GAS) based flexures to improve the acceleration sensitivity with low natural frequencies[1,2]. Such low frequency and soft spring devices are also becoming more useful for tilt sensing, and seismometer[3]. Low frequency devices, however, are sensitive to temperature changes, and hence the stability of the sensor is significantly impacted by any ambient temperature fluctuations[4]. Finite element analysis (FEA) is limited to estimate a poor performance of such devices. Thus, it needs to accurately characterize them particularly when any manually packaging steps are being adopted such as glued pickup glass plate on the MEMS gravimeter device. Temperature influenced effect is in part due to the negative linear temperature coefficient of Young's modulus (TCE) of ~ -63.83 ppm/°C [5–7] for silicon. A change in temperature leads to either the stiffening (when the temperature decreases) or softening (when the temperature increases) of the flexure springs that cause a corresponding displacement of the proof mass (PM). Assuming the designed frequency of the MEMS gravimeter is 7.5 Hz, the proof mass sags 5.5 µm with a 20 °C temperature increase, leading to a ~61.76 µGal/mK spurious change in the measured gravity signal. This thermal sag becomes significantly important if the gravimeter needs to measure a gravity change in the order of 10s of µGal, which is often the case for surveying applications in gravimetry. In order to compensate for the thermal sag, it is necessary to quantify the effect so that an effective compensation technique can be implemented. To remove the impact of temperature from the measurements, one can either monitor the temperature and regress it out during the post-processing step or implement passive or active compensation techniques. All of these approaches, however, have their advantages/disadvantages that need to be considered before implementation of such techniques. Considering that GAS based MEMS designs are gaining popularity in the recent years[8,9], it is crucial to understand the impact of temperature on such devices. Whilst there has been some previous work on formulizing the temperature sensitivity of GAS flexures used for test-mass suspensions in gravitational wave detectors[10,11], a complimentary study for the MEMS scale designs has not been adequately covered. Here, the thermal sag of the proof mass with an increase in temperature is quantified using the bespoke experimental setup, analytical and FEA. Various measurement techniques have been demonstrated to measure thermal sag in other devices. Kamp[12], has demonstrated the measurement of a vertical seismic accelerometer deflection using a laser Vibrometer with a 45° tilted mirror (a costlier technique). They have not, however, shown a temperature induced sag measurement. Imperial College London has used a laser source to measure the thermal sag of a vertical seismic accelerometer[13]. Such system needs a polished surface with a stable laser, however, and hence becomes a complex technique for thermal sag measurement.

In this paper we demonstrate a new approach with a simpler optical measurement system that measure the thermal sag of a MEMS gravimeter and verify the technique using both theoretical estimates and finite element analysis (FEA) simulations. The optical technique demonstrated in this paper is simpler and an alternative low-cost solution for thermal sag measurement. Such thermal measurements avoid the requirement of integrating a complete device using electronics and/or using more expensive instrumentation to be able to extract the thermal sensitivity of the device. Additionally, the current optical set-up has the potential to measure displacements under 1g (where g = 9.8066 ms$^{-2}$) loading in low-frequency devices before packaging and wire bonding. This information can also be used as a feedback to optimize the fabrication of devices. The rest of this paper is divided into the following sections: Section II contains an explanation of the analytical theory, Section III is an introduction to the thermal sag measurement setup used to conduct the work, Section IV is a description of the results and discussions, and Section V outlines the major conclusions and future directions.

## II. THEORY

A GAS MEMS gravimeter consists of

geometrical anti-spring based flexures, a proof mass and a frame as shown in Fig. 1. A photograph of the MEMS device (Fig 1a), the packaged MEMS gravimeter (Fig. 1b)[14], and a simplified schematic diagram of the GAS MEMS gravimeter are presented in **Error! Reference source not found.**. Fig. 1c also helps to visualize a thermal sag under the influence of temperature at an acceleration of 1g. The change in the deflection of the proof mass can be explained using the relation given in Eq. 1.

$$\Delta y_{eq} = \frac{\Delta F}{k_{eff}} = \frac{g \Delta m}{k_{eff}} = \frac{g \Delta m}{m \omega_0^2} \qquad (1)$$

where $\Delta y_{eq}$ is the thermal sag due to the change in temperature, $\Delta F$ is the change in applied effective load due to temperature, $m$ is the mass of the proof mass, $\Delta m$ is the change in the proof mass that is analogous to the change in Young's modulus of the material with temperature, $k_{eff}$ is the effective stiffness, and ω0 is the angular resonant frequency of the device. It can be also stated that:

$$\frac{\Delta y_{eq}}{\Delta T} = \left(\frac{g}{\omega_0^2}\right) \frac{\Delta E}{E \Delta T} \qquad (2)$$

A relative change, $\Delta E$, in the Young's modulus, E, due to a temperature change, $\Delta T$, is equivalent to a relative change of its load at fixed temperature: $\Delta E/E = \Delta m/m$ [11].

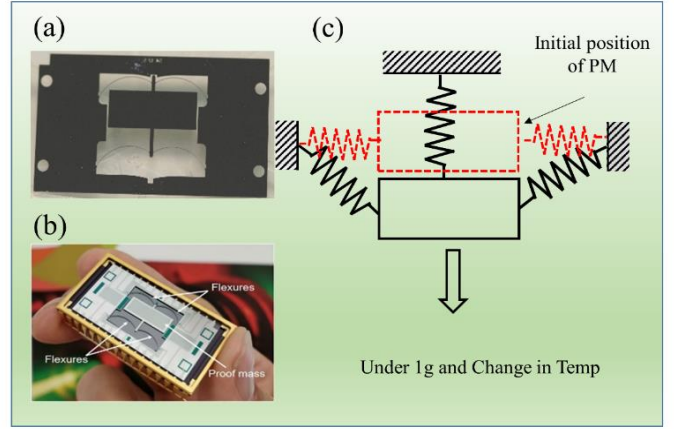

**Fig. 1**: (a) A photograph of the MEMS gravimeter used for the thermal sag measurements (its fabrication details can be accessed in [1]) (b) a photograph of the packaged MEMS gravimeter [12], and (c) a schematic diagram of the MEMS gravimeter under the influence of temperature.

As the tendency of silicon is to become soft under elevated temperature and it leads to the change in the stiffness of silicon material, which is responsible for the thermal sag of proof mass. The eq. 2 was used for analytical estimate of thermal sag vs frequency of device and estimate has the close agreement with FEA and experimental data which will be explained in the upcoming sections.

## III. THERMAL SAG EXPERIMENTAL SETUP

The MEMS gravimeter is operated vertically and hence measuring the position of the proof mass becomes difficult using the conventional available instruments inside a cleanroom during fabrication. In addition, temperature induced effects in the MEMS gravimeter are required to measure to understand the temperature sensitivity. It needs to increase the temperature in a controlled manner and measure the position of PM at different temperature. To address this issue, an optical measurement system was developed as shown in Fig. 2, this was based upon the work of Toland [15]. The bespoke experimental setup up consists of a Thorlabs high speed CMOS camera (Thorlabs DCC1240M) with 28x zoom optical lens (Thorlabs MVL12X20L, MVL20A and MVL12X12Z) in order to image the position of the PM of the MEMS gravimeter, a heater (Thorlabs PTC-1) (which can be controlled remotely to increase and decrease the temperature), and a cold light source (KL 1500 LCD) to illuminate the MEMS PM. The setup was kept on top of a vibration isolation bench to avoid any external coupled vibrations. Additionally, the MEMS, heater and camera were covered using a transparent, plastic box to avoid any issues associated with convection currents (Fig. 2).

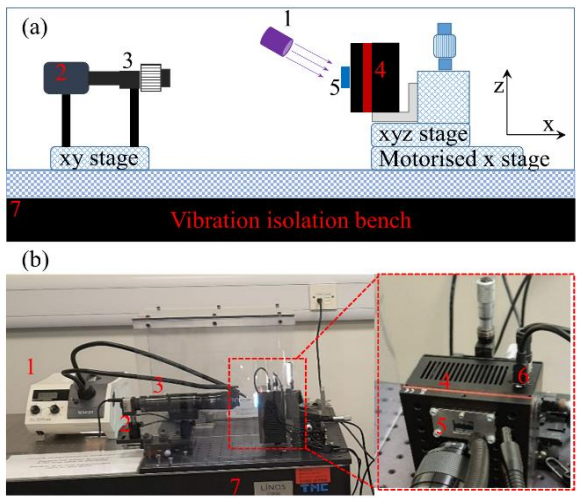

**Fig 2:** (a) a schematic and (b) a photograph of the thermal sag measurement setup with an expanded view of the MEMS device on a heater: (1) a cold light source, (2) CMOS camera, (3) 28x objective lens, (4) hot plate, (5) MEMS with a holding package, (6) to connect power supply and computer to hotplate and (7) the vibration isolation bench.

Remote accesses were used to record images using camera, to control the hotplate, and to move a motorized stage to adjust the focus. External disturbances from human interaction were therefore reduced.

In developing the setup, a few critical hurdles were noted that could reduce the efficacy of the apparatus. The foremost point was the thermal expansion coefficients of the materials used. During an initial test, a power resistor was utilized to increase the temperature of the MEMS, with the MEMS device placed on top of this power resistor using a slit made of two glass slides. The power resistor enclosure was made of aluminium, and due to the thermal expansion, the slit width was increased and hence, the MEMS was tilting. Due to the tilting, the effective gravity applied on the MEMS PM was reduced and it was observed that the proof mass was going up instead of down with an increase in the temperature. It can be concluded that the tilting effect was more significant compared to the softening of the silicon material. To remove the tilting issue of the MEMS device, a package was fabricated using stainless steel to hold the MEMS and it was kept on top of the hotplate instead of the power resistor.

After this change, it was possible to see the thermal sag using the optical experimental setup based on the differential measurements of the proof mass movement. The CMOS camera was first calibrated against a standard known diameter (25 μm) of wire and using a 28x optical zoom objective lens, the system had a calibration factor of 0.19 μm/pixel. The numerical aperture of the objective lens was 0.202 and working distance was 37 mm. The resolution of the CMOS camera was 1280 ×1024 pixels. As per the calibration factor, the measurement setup can image a 243 μm × 205 μm area of the MEMS device. During the course of the measurements at two different temperatures, it was observed that it was possible to move the full MEMS device up or down due the thermal expansion of the package and hotplate. Hence, a differential measurement is required to quantify the thermal sag. It was possible to see the frame and proof mass of the MEMS device simultaneously, and the artifacts spots on top of the MEMS device were utilized for differential measurements. The movement of the proof mass was recorded at 25˚C and 45˚C respectively and the artifacts were tracked using the Fiji ImageJ software [16]. The results of the thermal sag experiments and temperature sensitivity are presented in the next section.

## IV. RESULTS AND DISCUSSIONS

The MEMS gravimeter was characterized using the experimental setup as explained in Section III. The thermal sag measurements were performed remotely overnight on the vibration isolation bench to avoid any externally coupled vibrations. The movements of the proof mass at 25˚C and 45˚C were recorded and further analyzed to quantify the proof mass location using an open source ImageJ software for thermal sag of the MEMS gravimeter. Primarily, to understand measurement setup accuracy, the methodology used to for thermal sag measurement has been presented. Multiple images (frames) of the MEMS with moving PM and supporting frame were recorded. The photographs with actual and linearly changed pixel intensity are presented in Fig. 3 (a) and (b) respectively. Two points were selected on the MEMS, one on the frame as a reference and the second on the proof mass as a moving point. Both the points were tracked using the ImageJ software and proof mass position was calculated with respect to the reference point by taking difference.

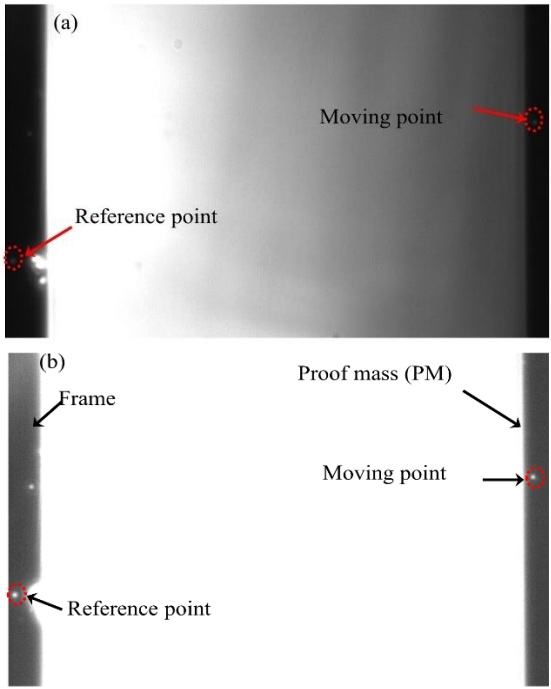

**Fig. 1**: (a) An actual and (b) linearly changed pixel intensity photographs for MEMS gravimeter having a moving PM (right) and a frame to attach the PM (left). Under the 1g loading, the proof mass continuously moves up and down because of ground vibration. The spot (red circled) on the left was used as a stationary reference and the spot (red circled) on the right was the moving point. The difference between these two points was then calculated

The difference between the two points was calculated and plotted against the respective frames imaged for PM movement shown in Fig. 4 (a) and a zoomed view for frames number 30 to 40 are presented in Fig. 4 (b) to see variation in the amplitude of PM movement. The variation in the amplitude is because of the ground vibrations. Further, the average position of PM with reference point was calculated and analysed. It is observed that a deviation of ±0.85 pixel (0.16 µm) was calculated in the measured average position of PM with reference position as shown in Fig. 4 (c). The differential measurements were conducted to avoid movement of the full MEMS device due the thermal expansion of the hot plate surface and/or the package used to hold the MEMS device. Fig. 4d shows the proof mass (PM) positions in pixels for temperature at 25 °C and 45 °C respectively and the difference in the position of proof mass was calculated based on the 0.19 µm/pixel calibration factor. Multiple MEMS gravimeter devices (named as MEMS 1, MEMS 2 MEMS 3, and MEMS 4) were characterized to locate the position of the proof mass with respect to the reference point at 25˚C and 45˚C.

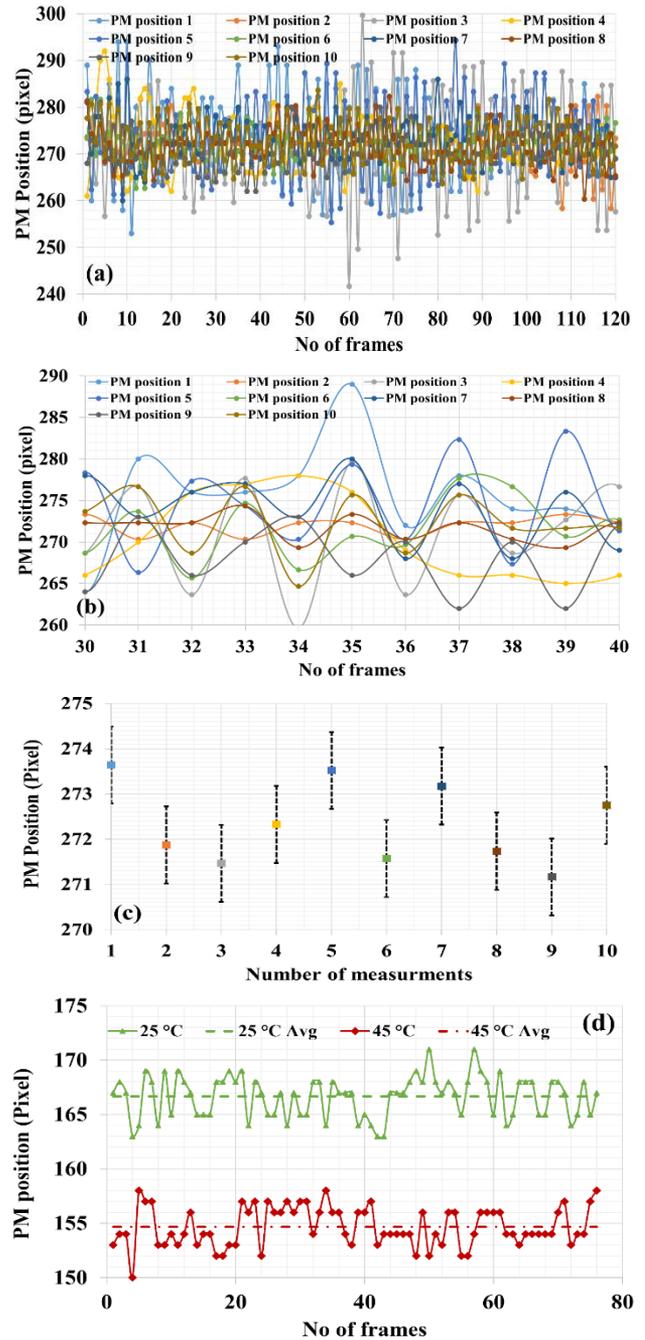

**Fig. 2:** (a) The 10 measurements for PM position recorded at 25˚C. (b) A zoomed view to see variation in the amplitude of PM movement for frame 30 to 40 for all measurements PM position 1 to 10. (c) An average position of the PM was calculated and plotted with the standard deviation (±0.85 pixel) in the PM position, (d) thermal sag measurement, green and red curves, shows the movement of the proof mass at 25 °C and 45 °C respectively. The image frames were recorded at 20 frames per second (FPS)

The thermal sag of a MEMS device was analytically calculated as a function of resonant frequency and presented in Fig. 5a by the black dashed curve. The MEMS devices investigated had frequencies ranging from of 10-12 Hz. The frequency was reduced to 8.5 Hz by adding mass to the proof mass so that the behaviour could be analysed for a lower frequency device. As can be seen in Fig. 5, the experimentally measured thermal sags were ~ 2 to 4 µm for devices under thermal sag investigation for a temperature change of 20 °C. A number of thermal sag experiments were conducted, the data were analysed and presented in Fig. 5 for all MEMS devices.

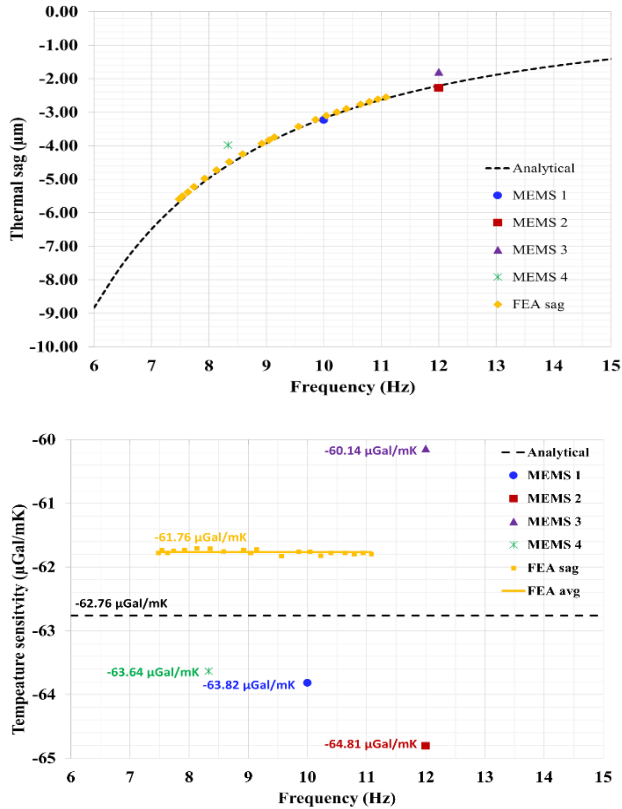

**Fig. 3**: (a) The thermal sag measurements and (b) temperature sensitivity of the MEMS gravimeter. The black dashed line shows an analytical estimate of the thermal sag of the MEMS gravimeter as a function of frequency. The yellow diamond points are thermal sag calculated using the FEA analysis. The solid blue circle, purple triangular and red square points are measured thermal sag by changing the temperature from 25°C to 45°C of the MEMS gravimeter. The thermal sag measurements were conducted for a 20°C change in the temperature.

It can be seen that the experimental thermal sag demonstrates a good agreement with the analytical calculations and the Ansys FEA simulation. For MEMS 1 (10 Hz device), the blue solid circle point shows the thermal sag measurements without the vibration isolation bench with measurements conducted during the day-time. Day-time experiments produced higher amplitudes of movement of the proof mass due to the externally coupled vibrations. Since differential measurement were carried out, these thermal sag experiments still show a good agreement with the estimated thermal sag. The issue with the externally coupled vibration was resolved by utilising the vibration isolation bench with overnight measurements (red square purple triangle points for MEMS 2 and 3 respectively of 12 Hz device). The MEMS 4 (8.5 Hz device was also characterized for the thermal sag (star green point) (Fig. 5a). Based on the all experiments, it can be concluded that the thermal sag can be measured and demonstrates good agreement with the estimated thermal sag with a deviation of 1.68 %, 3.26, 11.96 % and 9.56 % for MEMS 1, 2, 3 and 4 respectively. The temperature sensitivity was derived and plotted in the Fig. 5b and it can be seen that the temperature sensitivity of silicon based MEMS gravimeter was 60.14 - 64.81 µGal/mK, 61.76 µGal/mK and 62.76 µGal/mK for experimental, FEA simulation and analytical calculations respectively. It suggests that the temperature sensitivity is dependent on the inherent material properties used for fabrication of MEMS devices. If the silicon material properties can be tuned then it is possible to make the temperature in sensitive MEMS gravimeters.

## V. CONCLUSION

This paper demonstrates the performance of MEMS gravimeters under the influence of temperature. It demonstrates the significance of the thermal sag of the MEMS gravimeter on a device that can measure gravity changes of a few 10s of µGal. It has also shown a bespoke optical measurement setup to quantify the thermal sag of the MEMS devices. The measured results from the optical measurements have a very close agreement with the analytical and FEA calculations. Such a measurement system is required before any techniques to implement suitable compensation technique to mitigate the temperature induced effects of the device can be verified.

### ACKNOWLEDGE

This work is funded by the UK National Quantum Technology Hubs in Quantum Enhanced Imaging (EP/M01326X/1), Sensing and Timing (EP/T001046/1), NEWTON-g project funded by the EC's Horizon 2020 programme, under the FETOPEN-2016/2017 call (Grant Agreement No 801221) and Royal Academy of Engineering grants CiET2021_123 and RF/201819/18/83.